\documentclass[12pt,preprint]{aastex}

\slugcomment{AJ}

\shorttitle{X-ray and broad line luminosities in ULIRGs}
\shortauthors{Imanishi et al.}

\begin{document}

\title{X-ray Underluminous Active Galactic Nuclei Relative to Broad
Emission Lines in Ultraluminous Infrared Galaxies}  

\author{Masatoshi Imanishi \altaffilmark{1}}
\affil{National Astronomical Observatory, 2-21-1, Osawa, Mitaka, Tokyo
181-8588, Japan} 
\email{imanishi@optik.mtk.nao.ac.jp} 

\and

\author{Yuichi Terashima}
\affil{The Institute of Space and Astronautical Science, 3-1-1
Yoshinodai, Sagamihara, Kanagawa 226-8510, Japan}
\email{terasima@astro.isas.ac.jp} 

\altaffiltext{1}{Department of Astronomy, School of Science, Graduate 
University for Advanced Studies, Mitaka, Tokyo 181-8588}
   
\begin{abstract}
We present X-ray spectra of four ultraluminous infrared galaxies
(ULIRGs) with detectable broad near-infrared emission lines
produced by active galactic nuclei (AGNs): Mrk 463, PKS 1345+12,
IRAS 05189$-$2524, and IRAS 07598+6508. With the exception of IRAS 07598+6508,
high quality X-ray spectra obtained with {\it XMM} or {\it
Chandra} show modest (30--340 eV) equivalent widths of
the 6.4 keV Fe K$\alpha$ emission line and clear signatures for
absorption at a level of 4--33 $\times$ 10$^{22}$ cm$^{-2}$ for the main
power-law components from the AGNs. These spectral properties are
typical of Compton-thin AGNs, and so we estimate absorption-corrected 2--10
keV X-ray luminosities for the AGNs, Lx(2--10keV), using the
Compton-thin assumption. We compare the Lx(2--10 keV) values with
broad optical/near-infrared emission-line luminosities, and confirm a
previous finding by Imanishi \& Ueno that the Lx(2--10 keV) to
broad-emission-line luminosity ratios in ULIRGs are systematically lower
than those of moderately infrared-luminous type-1 AGNs. A comparison of 
independent energy diagnostic methods suggests that the AGNs are
underluminous in the 2--10 keV band with respect to their overall
spectral energy distributions, as opposed to the broad emission
lines being overluminous. This X-ray under-luminosity should be taken
into account when using 2--10 keV X-ray data to investigate the energetic
contribution from AGNs to 
the large infrared luminosities of ULIRGs.
\end{abstract}


\section{Introduction}

The star-formation rate is one of the most important indicators of the
nature of a galaxy; it is usually estimated from 
continuum or emission-line luminosities \citep{ken98}. In
dusty objects, the infrared luminosity is particularly important,
because the correction for dust extinction is generally not
significant. An observed infrared luminosity can be converted into a
star-formation rate, assuming that the luminosity is powered
primarily by star-formation activity. Although this assumption
seems to be true in the majority of moderately infrared-luminous
galaxies with L$_{\rm IR}$ $<$ 10$^{12}$L$_{\odot}$, it has been
suggested that active galactic nuclei (AGNs) provide an important 
energetic contribution in ultraluminous infrared galaxies (ULIRGs) with 
L$_{\rm IR}$ $>$ 10$^{12}$L$_{\odot}$ (Veilleux, Kim, \& Sanders 1999a).
Therefore, the AGN contribution to the observed infrared
luminosity of ULIRGs needs to be properly estimated.

To estimate this contribution, an indicator which can trace AGN
power, by disentangling from star-formation activity, is required. 
Two useful indicators are currently known. The first is the
absorption-corrected intrinsic 2--10 keV X-ray luminosity. 
In an AGN, strong 2--10 keV emission is produced by the inverse-Compton
process in the close vicinity of an accretion disk around 
a central supermassive black hole, whereas the 2--10 keV emission
from star-formation is usually much weaker. The second indicator is the
broad hydrogen emission lines in the optical and near-infrared domains,
whose line widths are larger than $\sim$1500 km s$^{-1}$ in full-width
at half maximum (FWHM). Such broad emission lines are believed to
originate in high-velocity gas (the so-called ``broad-line
regions'') around a central supermassive black hole in an AGN,
rather than being associated with phenomena related to star formation
(Veilleux, Sanders, \& Kim 1997b). In many dust-unobscured type-1 AGNs with
moderate infrared luminosity, both intrinsic 2--10 keV and
broad emission-line luminosities have been derived, and a correlation
between these luminosities has been found \citep{war88}. If this
correlation holds for all types of AGNs, then both of these indicators
can be used to estimate AGN's power in a galaxy.

However, results using {\it ASCA} data have shown that the
absorption-corrected 2--10 keV to broad emission-line
luminosity ratios in some ULIRGs are systematically lower than
those of dust-unobscured, moderately infrared-luminous,
type-1 AGNs \citep{ima99a}. One possible explanation is the large time
variability in X-ray luminosity. Some AGNs that were initially
classified as X-ray underluminous based on X-ray data taken at one
epoch showed normal X-ray luminosities when
observed at a later epoch \citep{ris03}. A second possibility is
the misidentification of the origin of the detected 2--10 keV emission. 
Based on modest equivalent widths of the Fe K$\alpha$ line
at 6.4 keV (EW$_{6.4}$ $<$ several $\times$ 100 eV) and signatures for
Compton-thin 
absorption (N$_{\rm H}$ $<$ 10$^{24}$ cm$^{-2}$) from {\it ASCA}
spectra, absorption-corrected 2--10 keV luminosities, Lx(2--10 keV),
were derived, assuming that the detected X-ray emission is a direct
component from an AGN behind Compton-thin absorbing material
\citep{awa91,kro94,ghi94}.  
If, instead, the detected X-ray emission is a scattered/reflected
component of AGN emission behind Compton-thick absorbing material
(N$_{\rm H}$ $>$ 10$^{24}$ cm$^{-2}$), the intrinsic 2--10 keV
luminosity will be larger than the estimate based on the
Compton-thin assumption. In the ULIRG, Superantennae,
although no significant Fe K$\alpha$ emission was seen in an {\it
ASCA} spectrum due to limited photon statistics
\citep{ima99b,pap00}, a strong Fe K$\alpha$ line (EW$_{6.4}$ $\sim$
1.4 keV) was reported in a higher quality spectrum obtained later
with {\it XMM} \citep{bra03a}. This demonstrated that the detected
2--10 keV emission is produced by a scattered/reflected component,
which was difficult to be recognized based on {\it ASCA} data.  
Higher quality {\it XMM} and {\it Chandra} data obtained at a different
epoch to the {\it ASCA} era are very useful to (1) investigate
whether the X-ray underluminosity of AGNs in ULIRGs found by
\citet{ima99a} is caused by X-ray time variability and (2)
better understand the origin of detected 2--10 keV emission using
higher photon statistics at 2--10 keV and, in particular, near 6.4 keV. 

In this paper, we present {\it XMM} and {\it Chandra} X-ray
spectra of ULIRGs that have detectable broad optical/near-infrared emission
lines, in order to investigate the 2--10 keV to broad-line luminosity
ratios. Throughout this paper, we adopt H$_{0}$ $=$ 75 km s$^{-1}$
Mpc$^{-1}$. Luminosities derived in previous literature
using different values of H$_{0}$ were re-calculated.
Lx(2--10 keV) denotes the absorption-corrected 2--10 keV luminosity of
an AGN, unless otherwise stated. 

\section{Targets}

Our targets are Mrk 463, PKS 1345+12, IRAS 05189$-$2524, and
IRAS 07598+6508. Table~\ref{tbl-1} gives
detailed information on these sources.

Mrk 463 has a double nucleus with a separation of $\sim$4 arcsec at
1.6 $\mu$m \citep{sur99} and is optically classified as a
Seyfert 2 \citep{san88}. The eastern nucleus (Mrk 463E) is active
and shows a broad (FWHM $>$ 1500 km s$^{-1}$) Pa$\beta$
($\lambda_{\rm rest}$ = 1.28 $\mu$m) emission line in the
near-infrared (Veilleux, Goodrich, \& Hill 1997a). Based on an
{\it ASCA} spectrum, \citet{uen96} estimated an EW$_{6.4}$ $<$ 700
eV, N$_{\rm H}$ $\sim$ 2 $\times$ 10$^{23}$ cm$^{-2}$, and found the
absorption-corrected 2--10 keV luminosity to be Lx(2--10 keV)
$\sim$ 4 $\times$ 10$^{42}$ ergs s$^{-1}$. 
Mrk 463E was diagnosed to be powered by an
AGN, with no detectable star-formation activity in the infrared
(Genzel et al. 1998; Veilleux, Sanders, \& Kim 1999b; Imanishi
2002). It was found to be underluminous in the 2--10 keV range compared
with the [OIII] and broad optical/near-infrared line emission  
\citep{uen96,ima99a}.

PKS 1345+12 is a double-nucleus source with a separation of $\sim$2
arcsec, with the western nucleus $\sim$3 times brighter at
2 $\mu$m than the eastern nucleus \citep{sur99}. It is
optically classified as a Seyfert 2 \citep{san88}, and shows a broad
Pa$\alpha$ ($\lambda_{\rm rest}$ = 1.88 $\mu$m) emission line in
the near-infrared \citep{vei97b}. \citet{ima99b} estimated
an EW$_{6.4}$ $<$ 500 eV, N$_{\rm H}$ = 1--4 $\times$ 10$^{22}$
cm$^{-2}$, and Lx(2--10 keV) $\sim$ 3 $\times$ 10$^{43}$ ergs
s$^{-1}$, based on an {\it ASCA} spectrum. This source is thought to be
powered by an AGN \citep{vei99b}. It was found to 
be underluminous in 2--10 keV X-rays, compared to the broad
optical/near-infrared emission-line luminosity \citep{ima99a}.

IRAS 05189$-$2524 is a single-nucleus source \citep{san88}, optically
classified as a Seyfert 2 \citep{vei99a}, and shows 
a broad Pa$\alpha$ emission line \citep{vei99b}. 
\citet{sev01} later reported the presence of broad components of 
Pa$\beta$ and Pa$\alpha$ emission lines, and argued that the broad
H$\alpha$ ($\lambda_{\rm rest}$ = 0.65 $\mu$m) component 
detected in the previous optical spectrum \citep{you96} is a transmitted
component, rather than a scattered/reflected component. 
This source was not included in \citet{ima99a}, but \citet{sev01} later
presented {\it ASCA} and {\it BeppoSAX} X-ray spectra which show
EW$_{6.4}$ $\sim$ 100--150 eV, N$_{\rm H}$ = 4--9 $\times$
10$^{22}$ cm$^{-2}$, and Lx(2--10 keV) $\sim$ 2 $\times$ 10$^{43}$
ergs s$^{-1}$. 
From the Lx(2--10 keV) value, \citet{sev01} argued that the AGN is not
powerful enough to account for the bulk of the infrared luminosity. 
However, some independent diagnostics in the infrared suggested that 
this source is dominated by an AGN \citep{vei99b,soi00,imd00}.  

IRAS 07598+6508 is a single-nucleus source, optically classified as a
Seyfert 1 \citep{san88}, shows a broad Pa$\alpha$ emission line
\citep{tan94}, and is a member of the class of broad absorption line
(BAL) quasars \citep{lip94,hin95}. 
No X-ray emission was detected by {\it ASCA} \citep{ris00}.

\section{Observations and Data Analysis}

\subsection{XMM}

Mrk 463, IRAS 05189$-$2524, and IRAS 07598+6508 were observed with
the EPIC cameras (PN, MOS1, and MOS2) on board {\it XMM}, in
full-frame mode. Observing details are shown in Table~\ref{tbl-1}.
Standard data analysis procedures were employed, using the {\it
XMM} Science Analysis System (SAS version 5.4), together with
standard software packages (FTOOLS 5.2 and XSPEC 11.2). Events
recorded during high-background time intervals were removed, and only
events corresponding to pattern 0--4 for PN and pattern 0--12 for
MOS were included in the analysis. The most recent calibration files
retrieved from the XMM webpage and response matrices created with SAS were
used for the data analysis.

The source spectra of Mrk 463 and IRAS 05189$-$2524 were extracted
from 30 arcsec radius circular regions, whereas for IRAS 07598+6508 a
22.5 arcsec circle was used. Background spectra were extracted 
from nearby source-free circular regions of 60 arcsec radius.
Spectra were binned so that each energy channel had more than 20
counts. Since the PN spectra have the best photon statistics,
they were used for primary estimation of important parameters.
However, to improve the statistics, final spectral fitting was
performed using a combination of PN, MOS1, and MOS2, keeping the
relative normalization free, because there is some uncertainty
about the relative flux calibration of the three cameras.

\subsection{Chandra}

PKS 1345+12 was observed with {\it Chandra}. Table~\ref{tbl-1}
summarizes the observing information. A 1/8 sub-array mode of the
ACIS-S3 back-illuminated CCD chip was used, with a 0.4-sec frame
time. The data were reprocessed with CIAO 2.2.1 and CALDB 2.17.
Only {\it ASCA} (``good'') grades 0, 2, 3, 4, and 6 were used in
the analysis.  High background intervals were removed.

A nuclear spectrum was extracted from a
circular region with a radius of 8 arcsec. A background spectrum
was extracted from an annular region around the nucleus, and
subtracted from the nuclear spectrum.  The count rate from the
nucleus was 0.057 counts s$^{-1}$, and the effect of pileup was
negligible.

\section{Results}

\begin{figure}
\plotone{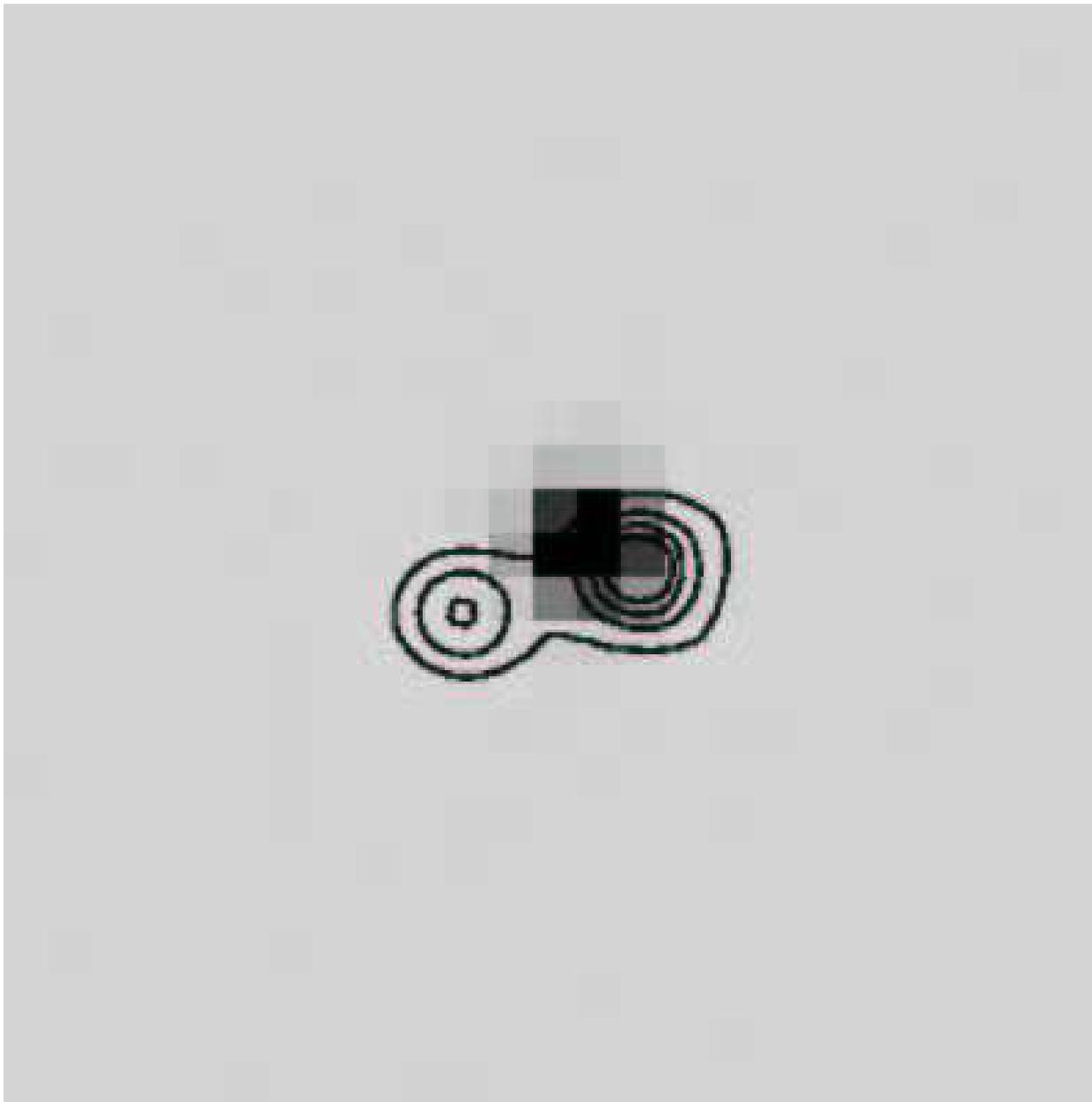}
\caption{
A {\it Chandra} 0.5--8 keV X-ray image of PKS 1345+12, with infrared
2.2-$\mu$m contours obtained from archival {\it HST} NICMOS data overplotted.  
North is up and east is to the left.  
The field of view is 12 $\times$ 12 arcsec$^{2}$.
The contours from the lowest one are 12.5\%, 25\%, 50\%, and 100\% of
the highest level at the western nucleus. 
The coordinate of the X-ray emission peak in this {\it Chandra} image is
J2000.0[13:47:33.37, 12:17:24.3], which is slightly displaced 
from the 2.2-$\mu$m peak of the western nucleus estimated from the NICMOS
archival data (J2000.0[13:47:33.33, 12:17:24.0]). 
\citet{eva99} made more accurate astrometry of this galaxy, and
estimated the coordinate of the 2.2 $\mu$m peak of the western
nucleus to be J2000.0[13:47:33.38, 12:17:24.4]. 
Considering the absolute positional uncertainty of {\it Chandra} (0.6
arcsec for 90\% error radius), we conclude that the X-ray emission comes
from the western nucleus.} 
\label{fig1}
\end{figure}

All sources are clearly detected in the {\it XMM} or {\it Chandra}
images. For PKS 1345+12, the high resolution {\it Chandra} image
reveals that the X-ray emission comes from the western nucleus
(Figure~\ref{fig1}). Figure~\ref{fig2} shows {\it XMM} EPIC
spectra of Mrk 463, IRAS 05189$-$2524, and IRAS 07598+6508, and a
{\it Chandra} spectrum of PKS 1345+12.

In the X-ray spectra of Mrk 463, PKS 1345+12, and IRAS
05189$-$2524, an absorbed hard component is clearly seen at $>$2
keV. This hard component is likely to originate in the AGN,
and so is fitted with an absorbed power law. In addition to this
hard component, there exists a soft X-ray component which dominates
the emission at $<$2 keV. There are two physically plausible
explanations for the origin of the soft X-ray emission: (1) a
scattered/reflected 
component of the primary AGN power-law emission, and (2) emission from a
starburst. 
In the caption of Table~\ref{tbl-2}, detailed descriptions of the models
used are reported for each source. 

For the spectral fitting of Mrk 463, a single temperature
thermal component (from a starburst) or a second power law 
(a Thomson-scattered component of the primary power law emission from
the AGN and/or X-ray binaries in a starburst) was initially included, in
addition to the main absorbed power-law component. However, significant
residuals remained at $<$2 keV. 
We then tried model (A), a low-temperature (0.5--1 keV) thermal model +
a high-temperature (5--10 keV) thermal component (from hot gas and X-ray
binaries originating in a starburst), and model (B), a
single-temperature (0.5--1 keV) thermal model + second power law. In
addition, since emission lines are seen at 6.5--7.0 keV, narrow
Gaussians at 6.4 keV, 6.7 keV, and 7.0 keV, K$\alpha$ emission lines
from Fe at different ionized levels, were also included in the model
fitting. Table~\ref{tbl-2} summarizes the best-fit models that were
adopted. The Lx(2--10 keV) depends only weakly on the adopted models. We
obtained EW$_{6.4}$ $\sim$ 340 eV, N$_{\rm H}$ $\sim$ 3 $\times$
10$^{23}$ cm$^{-2}$, and Lx(2--10 keV) $\sim$ 9 $\times$ 10$^{42}$ ergs
s$^{-1}$ in both models. 
The Lx(2--10 keV) is about twice as large as the estimate
based on {\it ASCA} data \citep{uen96}. The absorption-corrected
2--10 keV X-ray to observed infrared luminosity ratio is Lx(2--10
keV)/L$_{\rm IR}$ $\sim$ 5 $\times$ 10$^{-3}$.

For PKS 1345+12, a second power law and a narrow Gaussian (6.4 keV Fe
K$\alpha$ line) were included, in addition to the main absorbed
power-law component (model C). The X-ray 
spectrum is well fitted (Table~\ref{tbl-2}). We obtained
EW$_{6.4}$ $\sim$ 130 eV, N$_{\rm H}$ $\sim$ 5 $\times$ 10$^{22}$
cm$^{-2}$, and Lx(2--10 keV) $\sim$ 4 $\times$ 10$^{43}$ ergs
s$^{-1}$. The Lx(2--10 keV) we obtain is about 30\% higher than that derived
from {\it ASCA} data \citep{ima99b}. Lx(2--10 keV)/L$_{\rm IR}$ is
$\sim$6 $\times$ 10$^{-3}$.

For IRAS 05189$-$2524, model (D), a low temperature (0.5--1 keV)
thermal model + an even lower temperature ($<$0.3 keV) second thermal
component, and model (E), a single-temperature ($<$1 keV) thermal model
+ second power-law component, were used, primarily to account for the
soft X-ray 
emission. A narrow Gaussian for the Fe K$\alpha$ line at 6.4 keV was
also added, although this line is not as clear as in Mrk
463. Table~\ref{tbl-2} summarizes the adopted best-fit models. We
obtained EW$_{6.4}$ $\sim$ 30 eV, N$_{\rm H}$ $\sim$ 6 $\times$
10$^{22}$ cm$^{-2}$, and Lx(2--10 keV) $\sim$ 2 $\times$ 10$^{43}$
ergs s$^{-1}$. The Lx(2--10 keV) agrees with the {\it ASCA} and
{\it BeppoSAX} estimates \citep{sev01}. Lx(2--10 keV)/L$_{\rm IR}$
is $\sim$4 $\times$ 10$^{-3}$.

IRAS 07598+6508 shows no clear signature for an absorbed hard
component. The spectrum was fitted with a primary power law and a
narrow Gaussian for the Fe K$\alpha$ 6.4 keV line (Table~\ref{tbl-2}).
Although no strong Fe K$\alpha$ line is evident at 6.4 keV,
EW$_{6.4}$ could not be constrained strongly (EW$_{6.4}$ $<$ 13
keV), due to poor photon statistics. Lx(2--10 keV) and Lx(2--10
keV)/L$_{\rm IR}$ are estimated to be $\sim$8 $\times$ 10$^{41}$
ergs s$^{-1}$ and $\sim$9 $\times$ 10$^{-5}$, respectively.
BAL quasars are known to be weaker X-ray emitters than non-BAL quasars,
due to absorption \citep{gre01}. 
The substantially smaller Lx(2--10 keV)/L$_{\rm IR}$ in IRAS 07598+6508
(BAL quasar) than the other three ULIRGs may be related to the BAL
phenomena. 
The best fit value for the power law index $\Gamma$ $\sim$ 2.9 (Table 2)
is larger than the canonical value for AGNs ($\Gamma$ $\sim$ 1.8--2.0). 
The large $\Gamma$ value may be caused by the presence of additional
soft X-ray emission which can steepen the apparent X-ray spectrum. 
However, the limited photon statistics hamper more detailed
investigations of the spectrum.  

\section{Discussion}

\subsection{The 2--10 keV to Broad Line Luminosity Ratio}

For Mrk 463, PKS 1345+12, and IRAS 05189$-$2524, the modest
EW$_{6.4}$ ($<$ several $\times$ 100 eV) and clear signatures for
absorption with 
N$_{\rm H}$ $<$ several $\times$ 10$^{23}$ cm$^{-2}$ suggest that the
detected hard 
component is direct emission from an AGN behind Compton-thin
absorbing material \citep{awa91,kro94,ghi94}, and so the
absorption-corrected 2--10 keV luminosities, Lx(2--10 keV),
estimated in $\S$4, can be taken as the intrinsic X-ray luminosity of
the AGN. Therefore, the 2--10 keV 
to broad-line luminosity ratios for the AGNs in these ULIRGs can be
compared with those for moderately infrared-luminous type-1 AGNs
\citep{war88}. For IRAS 07598+6508, however, the constraint on 
EW$_{6.4}$ ($<$13 keV) is so loose that it is unclear whether the
observed 2--10 keV flux is a direct component behind Compton-thin
absorbing material (EW$_{6.4}$ $<$ several $\times$ 100 eV). 
The extremely small Lx(2--10 keV)/L$_{\rm IR}$ ratio and small 
N$_{\rm H}$ ($\sim$8 $\times$ 10$^{20}$ cm$^{-2}$) suggest that the
observed 2--10 keV emission is a scattered/reflected component behind
Compton-thick absorbing material.   
This source will be excluded from the following discussions.

The correlation between 2--10 keV and broad emission lines
established for type-1 AGNs \citep{war88} is for the broad
component of an optical H$\alpha$ ($\lambda_{\rm rest}$ = 0.65
$\mu$m) emission line. 
Thus, we need to convert the luminosity of a broad near-infrared 
emission line (Pa$\alpha$ for PKS 1345+12 and IRAS 05189$-$2524 
\footnote{
In IRAS 05189$-$2524, broad components were detected at H$\alpha$,
Pa$\beta$, and Pa$\alpha$ (see $\S$ 2). 
We use the Pa$\alpha$ line, because dust extinction is the lowest.   
}
; Veilleux et al. 1997b,1999b, and Pa$\beta$ for Mrk 463; Veilleux et
al. 1997a) 
to that of broad H$\alpha$.
For emission lines from the broad-line regions in an AGN, the case-B
assumption is not applicable to low-level transition lines, such as
H$\alpha$, due to collisional effects \citep{ost89}. 
For the broad H$\alpha$ to Pa$\alpha$ luminosity ratio, the value of 20
observed in dust-unobscured type-1 AGNs \citep{hil96} was adopted. 
Since the deviation of this ratio from case-B ($\sim$10) is only a
factor of $\sim$2, the choice of this ratio will not affect our main
conclusions significantly. 
For high-transition broad emission lines, such as Pa$\alpha$ and
Pa$\beta$, case-B is applicable \citep{rhe00}. 
Thus, a broad Pa$\alpha$ to Pa$\beta$ luminosity ratio of 1.9 (case-B)
was adopted. 

Absorption-{\it corrected} 2--10 keV and extinction-{\it
uncorrected} broad emission-line luminosities for the three ULIRGs
are plotted in Figure~\ref{fig3} (solid squares), where the estimated
broad H$\alpha$ emission-line luminosities are $\sim$5, $\sim$3, and
$\sim$1 $\times$ 10$^{43}$ ergs s$^{-1}$ for Mrk 463, PKS 1345+12,
and IRAS 05189$-$2524, respectively. The type-1 AGNs studied by
\citet{war88} are plotted for comparison (open circles). In addition to
the three ULIRGs, the other two Seyfert-2 ULIRGs with detectable broad
near-infrared Pa$\alpha$ emission lines, IRAS 20460+1925 and IRAS
23060+0505 \citep{vei97b}, have available {\it ASCA} 2--10 keV
spectra \citep{oga97,bra97}. Since their observed 2--10 keV 
fluxes are sufficiently high ($>$10$^{-12}$ ergs s$^{-1}$
cm$^{-2}$), strong constraints were obtained for EW$_{6.4}$
(260$^{+145}_{-137}$ eV for IRAS 20460+1925; Ogasaka et al. 1997,
$<$290 eV for IRAS 23060+0505; Brandt et al. 1997). The detected
2--10 keV fluxes show signatures for absorption with N$_{\rm
H}$ $=$ 2--8 $\times$ 10$^{22}$ cm$^{-2}$, so that the detected
2--10 keV flux is taken as a direct component from an AGN
behind Compton-thin absorbing material. For IRAS 20460+1925,
Lx(2--10 keV) is $\sim$1 $\times$ 10$^{44}$ ergs s$^{-1}$
\citep{oga97} and Lx(2--10 keV)/L$_{\rm IR}$ is $\sim$1 $\times$
10$^{-2}$. For IRAS 23060+0505, Lx(2--10 keV) is $\sim$2 $\times$
10$^{44}$ ergs s$^{-1}$ \citep{bra97} and Lx(2--10 keV)/L$_{\rm
IR}$ is $\sim$2 $\times$ 10$^{-2}$. Broad H$\alpha$ emission-line
luminosities were derived from Pa$\alpha$ data \citep{vei97b} using the
method described above and estimated to be $\sim$2 and 
$\sim$1 $\times$ 10$^{44}$ ergs s$^{-1}$ for IRAS 20460+1925 and
IRAS 23060+0505, respectively. These two ULIRGs are also plotted
in Fig.~\ref{fig3} (solid squares). The five Seyfert-2 ULIRGs show
absorption-{\it corrected} 2--10 keV to extinction-{\it uncorrected}
broad line luminosity ratios roughly an order of magnitude smaller than 
those in dust-unobscured type-1 AGNs with modest infrared
luminosities. 
While no significant dust extinction is expected for broad
emission lines in type-1 AGNs, some degree of dust extinction may be
present for the five Seyfert-2 ULIRGs.  
For these five ULIRGs, since near-infrared Pa$\alpha$ or Pa$\beta$
emission lines were used to estimate the broad H$\alpha$ luminosities,
the effects of dust extinction are expected to be reduced.  
If some amount of dust extinction is present for the broad near-infrared 
emission lines and its correction is applied, these five ULIRGs will
move to the right in Fig. 3, making the deviation of these ULIRGs even
larger.  
Even using new {\it XMM} and {\it Chandra} data, we confirmed
systematically smaller 2--10 keV to broad line luminosity ratios
in ULIRGs, originally suggested by \citet{ima99a}.

It is unlikely that the systematic difference is caused by the
uncertainty of the luminosity conversion from broad Pa$\alpha$ or
Pa$\beta$ to H$\alpha$ for the following two reasons. 
First, in a number of moderately infrared-luminous type-2 AGNs, the
2--10 keV to broad-H$\alpha$ luminosity ratios derived from observed
broad near-infrared line luminosities agree with those for unobscured
type-1 AGNs \citep{ima99a}.  
Second, for IRAS 05189$-$2524, the intrinsic broad H$\alpha$
luminosity estimated from the observed broad Pa$\alpha$ luminosity 
(L$_{\rm H \alpha}$ $\sim$ 1 $\times$ 10$^{43}$ ergs s$^{-1}$; see
previous paragraph) is in good agreement with that derived directly from
the broad H$\alpha$ luminosity \citep{you96} after correction for dust
extinction estimated by \citet{sev01} is applied (L$_{\rm H \alpha}$
$\sim$ 1 $\times$ 10$^{43}$ ergs s$^{-1}$). 
Both of these results suggest that (1) the luminosity conversion 
from broad near-infrared emission lines to broad H$\alpha$ is
reasonable, 
and that (2) at least for some sources with detectable broad
near-infrared emission lines, dust extinction is not significant in the
near-infrared domain.   
Thus, the significant deviation of ULIRGs in Fig. 3 is believed to be real.

\subsection{X-ray Underluminous or Broad Line Overluminous?}

Estimating the fractional AGN contribution to the
large infrared luminosities of ULIRGs is of particular importance to the
study of their energy sources. It is not clear which of the two AGN
indicators, 2--10 keV emission or broad optical/near-infrared
emission lines, is a better tracer of this fraction. The 2--10 keV 
luminosity depends primarily on the number of hot electrons in the close
vicinity of an accretion disk around a central supermassive blackhole
in an AGN and the number of incident UV 
photons to be Compton up-scattered. The broad emission-line
luminosity is roughly proportional to the product of the number of
UV photons from the central AGN and the covering factor of broad-line
gas clouds. The infrared luminosity of a ULIRG powered by an 
AGN is determined mainly by the number of UV photons from the AGN and by
the covering factor of dust. In the nuclear region of a 
ULIRG, a large amount of gas and dust is available \citep{san96},
which may increase the covering factors both of dust and broad-line gas
clouds around the AGN. If this is the case, broad emission lines are a
better indicator of the AGN's contribution to the infrared luminosity,
and AGNs in ULIRGs are underluminous in 2--10 keV X-rays. 

In fact, in a number of ULIRGs, although an AGN is consistently
suggested to be the primary energy source by some independent
methods, only energy diagnostics based on the 2--10 keV luminosity
suggest that the AGN is energetically insufficient. Examples 
include IRAS 05189$-$2524 \citep{vei99b,soi00,imd00,sev01} and Mrk
463 \citep{uen96,gen98,ima02}. Thus, it is more likely that the AGNs in
ULIRGs are underluminous in X-rays with respect to their overall
spectral energy distributions, than that their broad lines are
overluminous. 

If X-ray underluminosity (relative to UV) is an intrinsic property of
ULIRGs, one possible explanation is a high mass accretion rate onto
a central supermassive black hole \citep{bec03}. This is plausible in
ULIRGs, given the high nuclear concentration of gas and dust \citep{san96}.

Alternatively, the apparent X-ray underluminosity of AGNs in ULIRGs
may be caused by our misidentification of the origin of the
detected 2--10 keV emission from ULIRGs. Using reasonable
assumptions for an AGN, the scattered/reflected X-ray component of
AGN emission behind Compton-thick absorbing material is
expected to show a very large EW$_{6.4}$ with $>$1 keV
and only weak absorption \citep{mat03}. A modest EW$_{6.4}$ and
clear Compton-thin absorption signature, as were found in the five
ULIRGs, make it most plausible that we are observing direct X-ray
emission from an AGN behind Compton-thin absorbing material
\citep{awa91,kro94,ghi94}. 
However, Thomson scattering by highly ionized gas produces a small
EW$_{6.4}$. If the scattered component suffers some degree of
absorption, both the modest EW$_{6.4}$ and clear signature for
Compton-thin absorption can be explained \citep{mal00}. In this
scenario, the intrinsic 2--10 keV luminosity of an AGN
could be substantially larger, depending on the scattering
efficiency. Since such X-ray spectra are usually interpreted
as Compton-thin, and the absorption-corrected 2--10 keV
luminosity is determined based on this assumption, we will
underestimate the true Lx(2--10 keV).

A third possibility which could explain the observed X-ray
underluminosity is partial Compton-thick X-ray absorption. If the
X-ray-absorbing material in front of an AGN consists of small clumps of 
Compton-thick and thin clouds, whose size scales are substantially
smaller than the size of the X-ray-emitting region of the AGN, then
the 2--10 keV spectrum will suggest Compton-thin absorption. 
The actual Lx(2--10 keV) could be significantly larger than a simple
estimate assuming uniform Compton-thin X-ray absorption. The covering
factor of Compton-thick clouds in front of an AGN along our line-of-sight
determines the degree of our underestimate for Lx(2--10 keV), but
an X-ray spectrum at $>$10 keV is required to determine this \citep{bra03b}.

Whether the X-ray underluminosity of AGNs in ULIRGs is intrinsic or
caused by our underestimate of Lx(2--10 keV), the use of 2--10 keV
observations to estimate AGNs' contribution to the infrared luminosities
of ULIRGs will be quite misleading, unless the 
X-ray underluminosity is properly taken into account.

The discovery of underluminous X-ray emission relative to broad
emission lines from AGNs is only for ULIRGs with modest dust
obscuration along our sightlines, because in order for broad
optical/near-infrared emission lines to be detected, dust obscuration
toward AGNs must be relatively 
low. The putative AGNs in the majority of ULIRGs are so highly
dust-obscured that broad emission lines are not detectable
\citep{vei99b}, and so we have no information on the broad line 
luminosities. It is unclear whether X-ray underluminosity is a general
property of AGNs in the majority of ULIRGs, and there is currently 
no way to determine this. However, when using 2--10 keV data to
determine the energy contribution of AGNs in ULIRGs, it is important to
consider that the AGN contribution may be significantly underestimated. 

\section{Summary}

We estimated the absorption-corrected 2--10 keV X-ray luminosities
of AGNs in moderately dust-obscured ULIRGs, and confirmed
a previous finding that absorption-{\it corrected} 2--10 keV 
to extinction-{\it uncorrected} broad emission-line luminosity
ratios in ULIRGs are roughly an order of magnitude smaller than
those of type-1 AGNs with modest infrared
luminosities. We have shown that the smaller ratios are more likely to
be due to 2--10 keV X-ray underluminosity with respect to the overall
spectral energy distributions, rather than higher
broad-emission-line luminosity. The lower X-ray luminosities of AGNs in
ULIRGs should be taken into account in estimating AGNs'
contribution to the energy output from ULIRGs based on X-ray
data.

\acknowledgments

Y.T. acknowledges the Research Fellowship of the Japan Society
for the Promotion of Science for Young Scientists.
This work is based on observations obtained with XMM-Newton, an ESA
science mission with instruments and contributions directly funded by
ESA Member States and the USA (NASA), and with Chandra.
This research has made use of the SIMBAD database, operated at CDS,
Strasbourg, France, and of the NASA/IPAC Extragalactic Database 
(NED) which is operated by the Jet Propulsion Laboratory, California
Institute of Technology, under contract with the National Aeronautics
and Space Administration.

\clearpage

\begin{deluxetable}{lcclll}
\tabletypesize{\small}
\tablenum{1}
\tablecaption{Details of Observed ULIRGs. \label{tbl-1}} 
\tablewidth{0pt}
\tablehead{
\colhead{Object} & \colhead{z} &  \colhead{L$_{\rm IR}$} & 
\colhead{Satellite} & \colhead{Date} & \colhead{Net Exposure} \\
\colhead{} & \colhead{} &  \colhead{[ergs s$^{-1}$]} & 
\colhead{} & \colhead{[UT]}  & 
\colhead{[ksec]} \\
\colhead{(1)} & \colhead{(2)} & \colhead{(3)} & \colhead{(4)} &
\colhead{(5)} & \colhead{(6)}  
}
\startdata
Mrk 463  & 0.051 & 45.3 \tablenotemark{1} & XMM & 2001 Dec 22 & 21 (PN),
25 (MOS) \\ 
PKS 1345+12 & 0.122 & 45.8 & Chandra & 2000 Feb 24 & 19 \\
IRAS 05189$-$2524 & 0.042 & 45.7 & XMM & 2001 Mar 17 & 7 (PN), 10 (MOS)  \\
IRAS 07598+6508 & 0.149 & 46.0 & XMM & 2001 Oct 25 & 13 (PN), 19 (MOS) \\
\enddata

\tablecomments{
Column (1): Object. 
Column (2): Redshift. 
Column (3):  Logarithm of infrared (8$-$1000 $\mu$m) luminosity
in ergs s$^{-1}$ calculated with
$L_{\rm IR} = 2.1 \times 10^{39} \times$ D(Mpc)$^{2}$
$\times$ (13.48 $\times$ $f_{12}$ + 5.16 $\times$ $f_{25}$ +
$2.58 \times f_{60} + f_{100}$) ergs s$^{-1}$
\citep{san96}, where f$_{12}$, f$_{25}$, f$_{60}$, and f$_{100}$ are 
{\it IRAS FSC} fluxes at 12$\mu$m, 25$\mu$m, 60$\mu$m, and 100$\mu$m,
respectively.  
Column (4): X-ray satellite used for the observation.
Column (5): Observing date in UT. 
Column (6): Effective exposure time in ksec after discarding high
background intervals. 
}

\tablenotetext{1}{Mrk 463 is not a ULIRG in the strict sense, but is
included, since its infrared luminosity is close to that of a
ULIRG.}

\end{deluxetable}

\clearpage

\begin{deluxetable}{lccccc}
\tabletypesize{\small}
\tablenum{2}
\tablecaption{Results of X-ray Spectral Fitting. \label{tbl-2}}
\tablewidth{0pt}
\tablehead{
\colhead{Object} & \colhead{Model\tablenotemark{1}} & \colhead{$\chi^{2}$/d.o.f.} & 
\colhead{N$_{\rm H}$} & 
\colhead{L$_{\rm X}$(2--10keV)} & \colhead{EW$_{6.4}$} \\
\colhead{} & \colhead{} & \colhead{} & \colhead{[10$^{22}$ cm$^{-2}$]} &
\colhead{[10$^{42}$ ergs s$^{-1}$]} & \colhead{[eV]}  \\ 
\colhead{(1)} & \colhead{(2)} & \colhead{(3)} & \colhead{(4)} &
\colhead{(5)} & \colhead{(6)} 
}
\startdata
Mrk 463 & A & 231.6/202 & 33$^{+3}_{-6}$ & 9.2 & 340$^{+70}_{-100}$ \\ 
        & B & 230.4/203 & 32$\pm$3  & 9.1 & 340$^{+70}_{-90}$ \\ 
PKS 1345+12 & C & 49.2/62 & 4.5$^{+0.5}_{-0.4}$ & 38 & 130$\pm$130 \\ 
IRAS 05189$-$2524 & D & 225.9/232 & 5.7$\pm$0.3 & 17 & 30$^{+50}_{-30}$ \\ 
                  & E & 233.2/233 & 5.8$\pm$0.4 & 16 & 30$^{+50}_{-30}$ \\ 
IRAS 07598+6508 & F & 25.1/23 & 0.08$^{+0.10}_{-0.06}$ & 0.8 & $<$13000 \\ 
\enddata

\vspace{1cm}

Note. --- 
Column (1): Object.  
Column (2): Adopted best-fit models. 
Column (3): Reduced $\chi^{2}$ values. 
Column (4): Absorption for the main power law component.
The uncertainty is at 90\%
confidence level for one parameter of interest ($\Delta\chi^{2}$ =
2.7) throughout this Table. Column (5): Absorption-corrected 2--10
keV X-ray luminosity for the main power law component, estimated based on the assumption of uniform Compton-thin
absorption. Column (6): Equivalent width of the 6.4 keV Fe
K$\alpha$ emission line relative to the main power law component.

\tablenotetext{1}{
\noindent
Model A : absorbed power law ($\Gamma$ = 1.8 fixed) + 
  thermal (kT = 0.6$\pm$0.1 keV) + thermal (kT = 10 keV fixed) 
  + narrow Gaussian (6.4 keV Fe K$\alpha$) + 
  narrow Gaussian (6.7 keV, EW = 30$^{+50}_{-30}$ eV) + 
  narrow Gaussian (7.0 keV, EW = 120$^{+100}_{-70}$ eV) \\
\newline
Model B : absorbed power law ($\Gamma$ = 1.8 fixed) + 
  thermal (kT = 0.6$\pm$0.1 keV) +  
  second power law ($\Gamma$ = 1.8 fixed; $\sim$5\% of the main power law
  component in normalization) + 
  narrow Gaussian (6.4 keV Fe K$\alpha$) + 
  narrow Gaussian (6.7 keV, EW = 40$^{+50}_{-40}$ eV) + 
  narrow Gaussian (7.0 keV, EW = 130$\pm$80 eV) \\
\newline
Model C : absorbed power law ($\Gamma$ = 1.8 fixed) +  
  second power law ($\Gamma$ = 1.8 fixed; $\sim$3\% of the main power law
  component in normalization) + 
  narrow Gaussian (6.4 keV Fe K$\alpha$) \\
\newline
Model D : absorbed power law ($\Gamma$ = 1.8 fixed) + thermal 
  (kT = 0.7$^{+0.1}_{-0.2}$ keV) + thermal (kT = 0.11$\pm$0.02 keV) +
  narrow Gaussian (6.4 keV Fe K$\alpha$)  \\
\newline
Model E : absorbed power law ($\Gamma$ = 1.8 fixed) + thermal 
  (kT = 0.3$\pm$0.1 keV) + second power law ($\Gamma$ = 1.8 fixed;
  $\sim$1\% of the main power law component in normalization) + 
  narrow Gaussian (6.4 keV Fe K$\alpha$)  \\
\newline
Model F : absorbed power law ($\Gamma$ = 2.9$^{+0.6}_{-0.5}$) + 
  narrow Gaussian (6.4 keV Fe K$\alpha$)  \\
}

\end{deluxetable}

\clearpage

\clearpage 

\begin{figure}
\plottwo{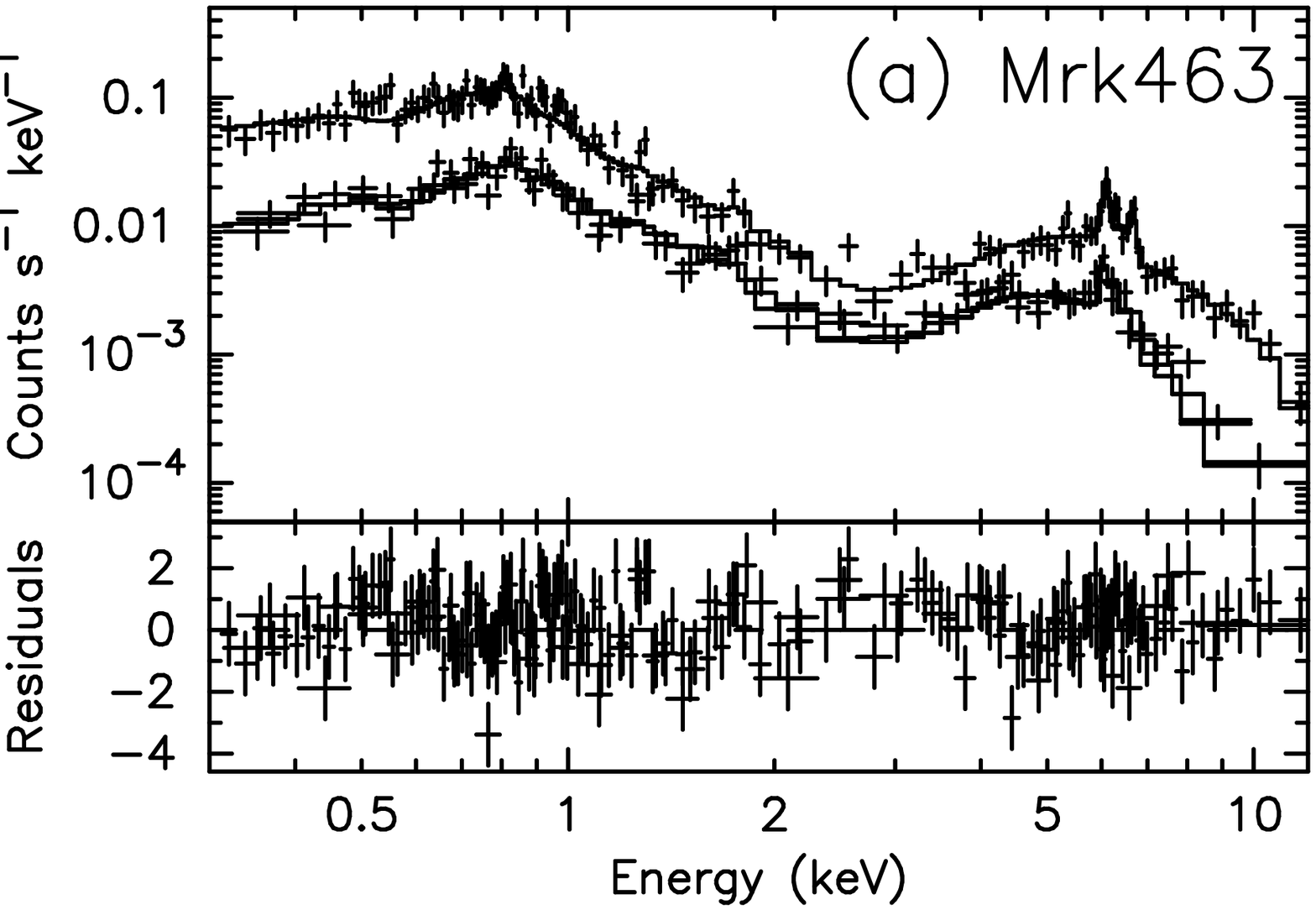}{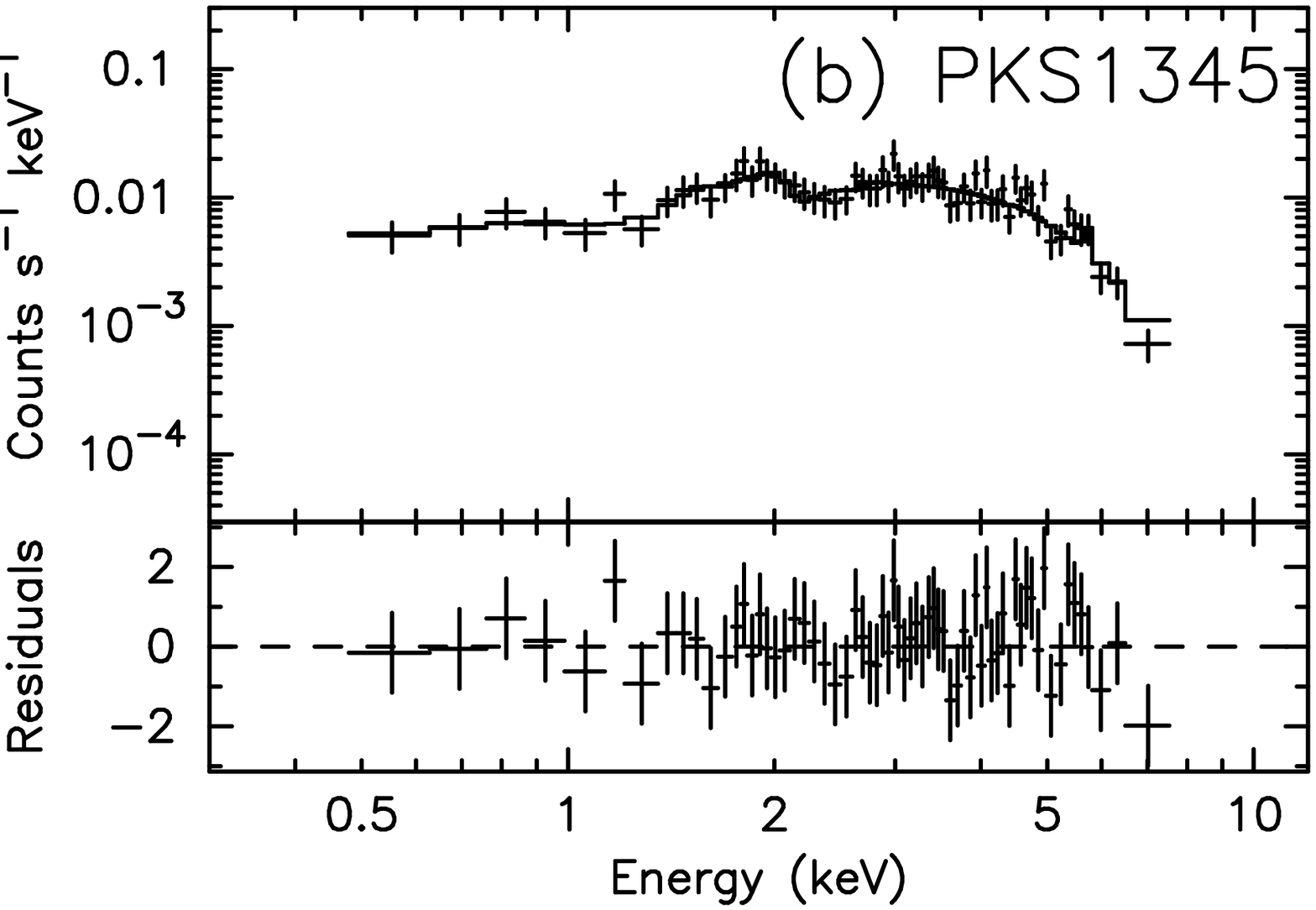}
\end{figure}
\begin{figure}
\plottwo{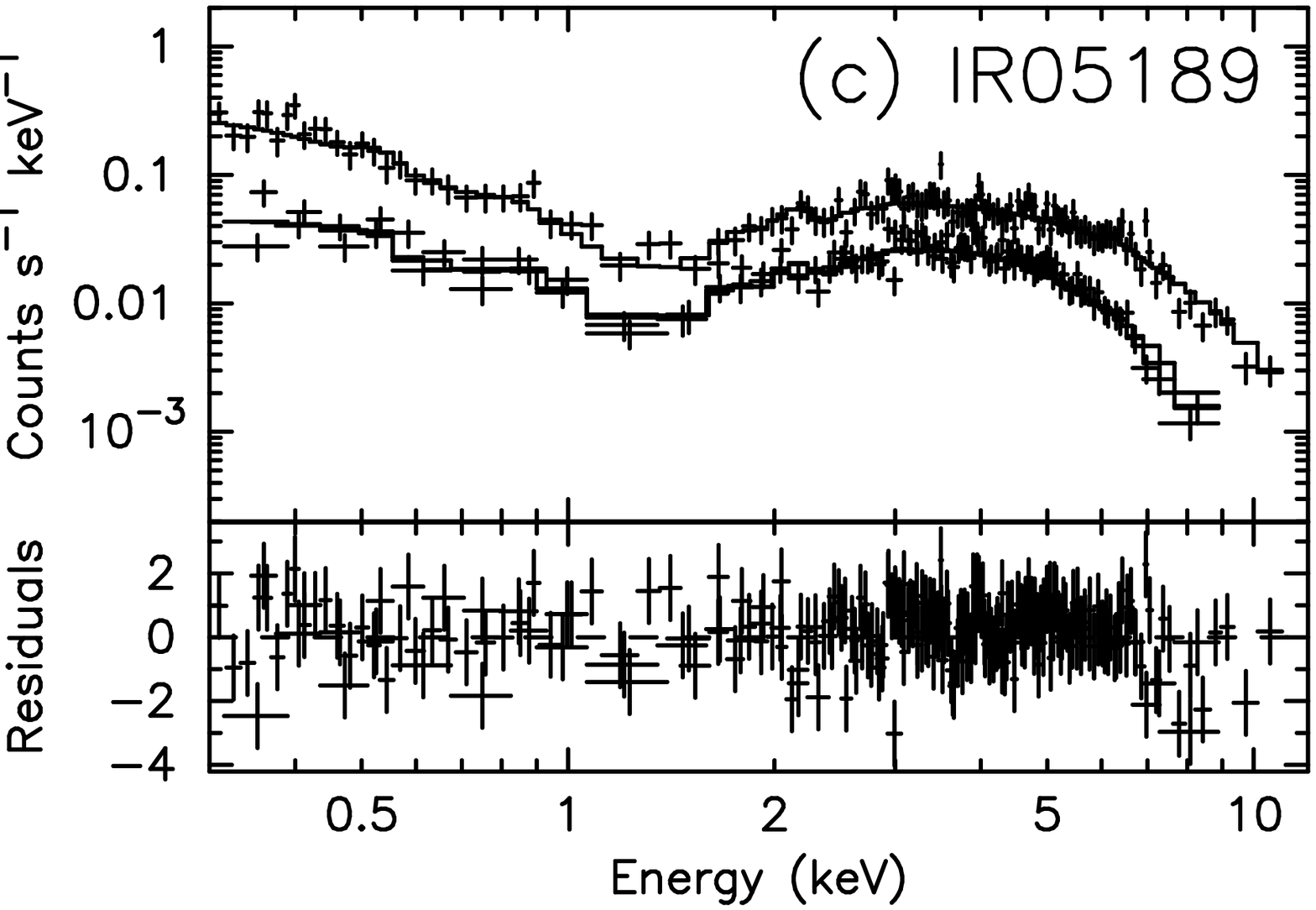}{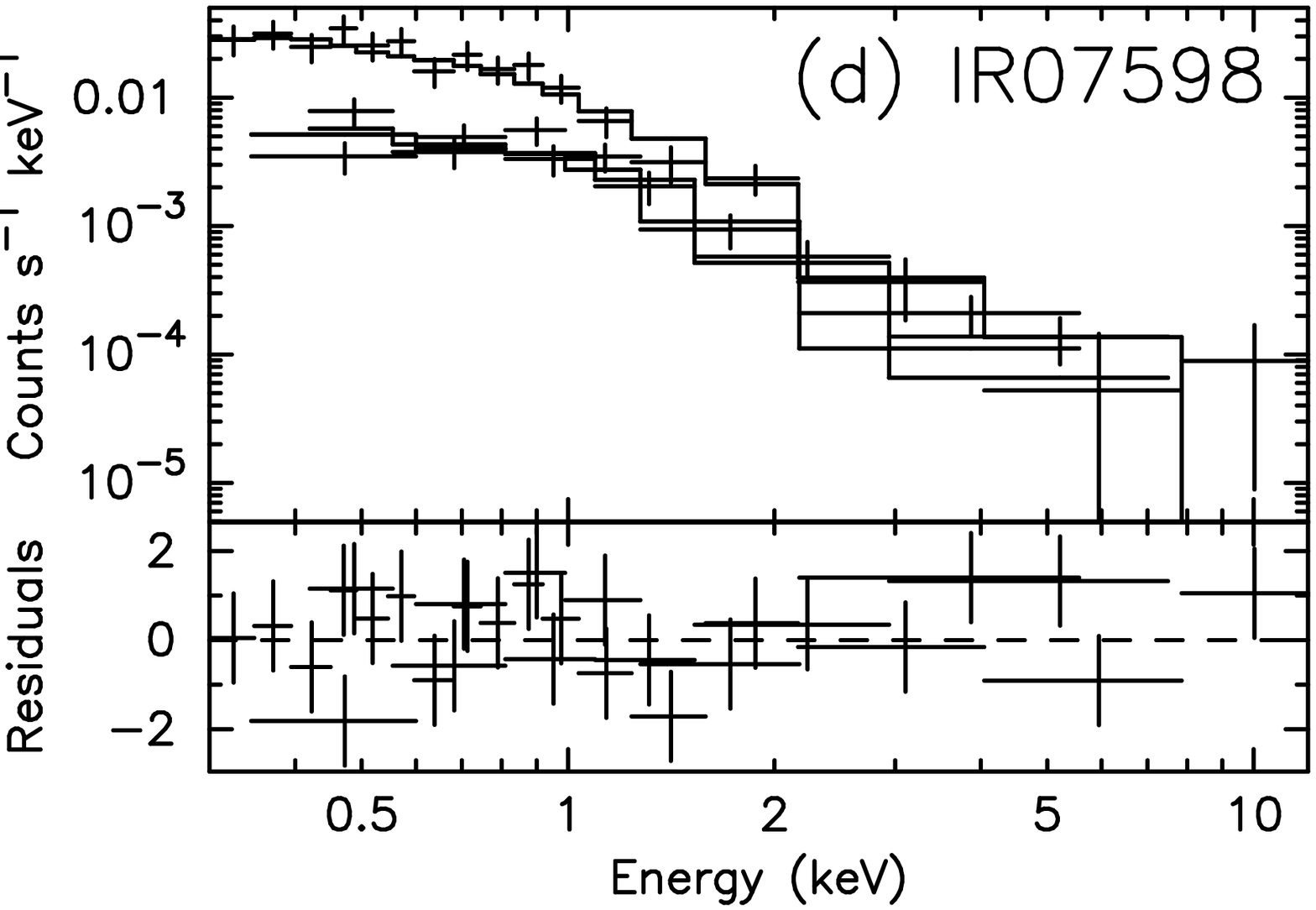}
\caption{
{\it (a)}: An {\it XMM} spectrum of Mrk 463. Model A is overplotted as
solid lines.
In the upper panel, the higher plots are a PN spectrum (which contains
the most counts), and the lower plots are MOS spectra.  
The lower panel shows residuals of data from the model. 
{\it (b)}: A {\it Chandra} spectrum of PKS 1345+12. Model C is overplotted.
{\it (c)}: An {\it XMM} spectrum of IRAS 05189$-$2524. Model D is overplotted.
{\it (d)}: An {\it XMM} spectrum of IRAS 07598+6508. Model F is overplotted.}
\label{fig2}
\end{figure}

\clearpage

\begin{figure}
\plotone{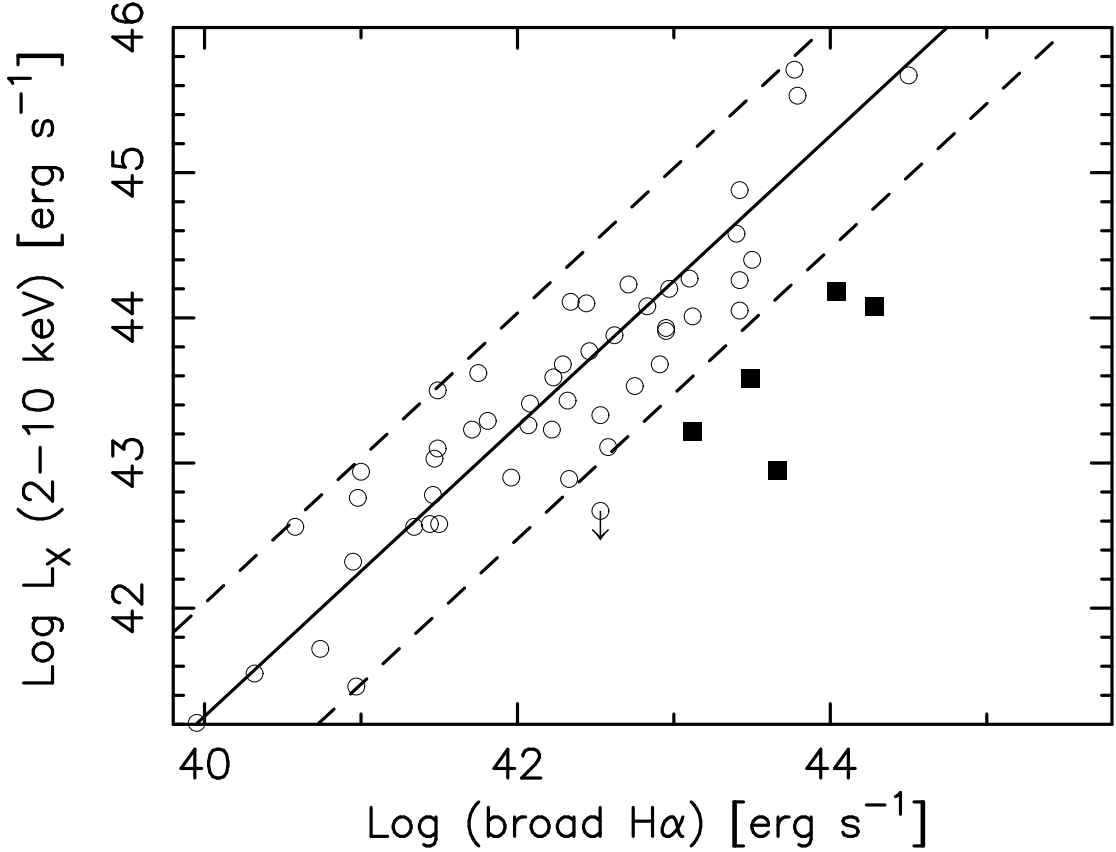}
\caption{
Absorption-{\it corrected} 2--10 keV X-ray luminosity
(ordinate) and extinction-{\it uncorrected} broad H$\alpha$
emission line luminosity (abscissa). Open circles are
dust-unobscured type-1 AGNs with moderate infrared luminosity
taken from \citet{war88}. Filled squares are the five Seyfert-2
ULIRGs studied in this paper, Mrk 463, PKS 1345+12, IRAS
05189$-$2524, IRAS 20460+1925, and IRAS 23060+0505 (see text). The
solid line is for log Lx(2--10 keV) = log L(broad H$\alpha$) +
1.255 \citep{ima99a}. In the dashed lines, Lx(2--10
keV)-to-L(broad H$\alpha$) ratio is a factor of 6 larger and
smaller than the solid line. The only X-ray underluminous AGN (IRAS
05218$-$1212) originally found by \citet{war88} is infrared selected by
{\it IRAS}, if not a ULIRG. Since no information on EW$_{6.4}$ is
available, the weak 2--10 keV X-ray emission might be caused by
Compton-thick absorption.} 
\label{fig3}
\end{figure}

\end{document}